# Acoustic type-II Weyl nodes from stacking dimerized chains

Zhaoju Yang[1] and Baile Zhang[1,2*]

[1]Division of Physics and Applied Physics, School of Physical and Mathematical Sciences;

[2]Centre for Disruptive Photonic Technologies,

Nanyang Technological University, Singapore 637371, Singapore.

*To whom correspondence should be addressed:

E-mail: blzhang@ntu.edu.sg

**Lorentz-violating type-II Weyl fermions, which were missed in Weyl's prediction of nowadays classified type-I Weyl fermions in quantum field theory, have recently been proposed in condensed matter systems. The semimetals hosting type-II Weyl fermions offer a rare platform for realizing many exotic physical phenomena that are different from type-I Weyl systems. Here we construct the acoustic version of type-II Weyl Hamiltonian by stacking one-dimensional dimerized chains of acoustic resonators. This acoustic type-II Weyl system exhibits distinct features in finite density of states and unique transport properties of Fermi-arc-like surface states. In a certain momentum space direction, the velocity of these surface states are determined by the tilting direction of the type-II Weyl nodes, rather than the chirality dictated by the Chern number. Our study also provides an approach of constructing acoustic topological phases at different dimensions with the same building blocks.**

Weyl fermions, originally predicted by Weyl [1] in 1929, have recently been discovered in Weyl semimetals [2-4], a new state of matter whose surface states form Fermi arcs linking Weyl nodes in the momentum space. In parallel, a double-gyroid photonic crystal [5] that exhibits Weyl nodes for electromagnetic waves has been demonstrated [6], starting a new chapter in photonics research [7]. Following these developments, Weyl nodes for acoustic waves have also been proposed by applying on-site unequal coupling or chiral coupling on a graphite structure [8]. Both coupling approaches can enable the emergence of acoustic waves as Weyl quasiparticles in an acoustic crystal.

Recently, type-II Weyl semimetals hosting fundamentally new type-II Weyl fermions have been proposed [9] and confirmed experimentally [10-13] with both the bulk fermions and Fermi arcs observed. A type-II Weyl node appears at the contact of electron and hole pockets and exhibits a strongly tilted cone spectrum with non-vanishing density of states (DOS), in contrast to a point-like Fermi surface at a type-I Weyl node [2] with vanishing DOS. The type-II Weyl fermions were in fact "missed" by Weyl because they violate the Lorentz symmetry. Therefore unlike the type-I case, the type-II Weyl fermions cannot be adiabatically connected back to the Lorentz-invariant Weyl fermions in Weyl's prediction.

In the context of acoustics, the previous proposal based on the graphite structure [8] did not distinguish type-I and type-II Weyl Hamiltonians. Consequently, only acoustic topological surface states for type-I Weyl nodes have been studied. In this Letter, on a platform of stacked dimerized chains of acoustic resonators, we construct acoustic type-II Weyl nodes following the explicit type-II Weyl Hamiltonian [9]. Unique features of this acoustic type-II Weyl system include the distinct finite DOS and transport properties of topological surface states. In a certain momentum space direction, the bands of the surface states have the same sign of velocity, which is determined by the tilting direction of type-II Weyl nodes rather than their chirality dictated by the Chern number. Because of the existence of an incomplete bandgap, the acoustic waves of the surface states can be scattered by defects and penetrate into the bulk, and thus do not exhibit the same robust propagation as demonstrated in Ref. [8].

We follow the line of thought in the previous proposal [8], but adopt 1D resonator chains as building blocks. Firstly, since Weyl nodes are three-dimensional (3D) extensions of two-dimensional (2D) Dirac nodes [5-7], we first construct 2D Dirac nodes by stacking 1D chains. Secondly, in view of the difficulty of achieving *T*-symmetry breaking in topological acoustics [14-16], we comply with the principle of *P*-symmetry breaking when stacking 1D resonator chains. Note that a 1D acoustic topological phase has been realized in a 1D phononic crystal [17], but not with resonators. It is thus also interesting to investigate if a 1D resonator chain can also realize the 1D acoustic topological phase.

We start to design the 1D dimerized chain of acoustic resonators following the Su-Schrieffer-Heeger (SSH) model [18], which is shown in the upper part of Fig. 1(a). The filled (open) circles indicate A (B) type atoms. The left and right nearest-neighbor (NN) hopping strengths with respect to an A type atom are $t+\delta t$ and $t-\delta t$, respectively. By setting zero energy offset between two sites, we can obtain the following Bloch Hamiltonian:

$$H_1(k) = 2t\cos(k_x a)\sigma_x - 2\delta t \sin(k_x a)\sigma_y, \tag{1}$$

where $\sigma_x$ and $\sigma_y$ are Pauli matrices. As shown in the lower part of Fig. 1(a), we construct one unit-cell of the chain with two identical resonators, connected by two coupling waveguides with different radii. The periodic boundary condition is applied to the left and right surfaces. Other surfaces (blue) are treated as sound hard boundaries. The distance between two nearest resonators is $a = 0.1$ m. The radius and height of each resonator (cylinder) is $r = 0.4a$ and $h = 0.8a$. For dimerization, we apply modulation of $\delta w = 0.3w$ to the original radius of coupling waveguide $w = 0.26r$. We thus have $w+\delta w$ ($w-\delta w$) for the left (right) coupling waveguide.

Since there are two resonators in one unit-cell, we only consider the two-band model with two lowest acoustic eigen-modes, whose acoustic fields simulated by finite-element commercial software COMSOL Multiphysics are nearly single-valued for each resonator. By choosing modulation $\delta w = 0.3w, 0, -0.3w$, we numerically calculate three band structures as shown in Fig. 1(b). The closing of bandgap at $\delta w = 0$ indicates the topological phase transition. For the lower bands of left and right parts, we can characterize their topological properties by calculating the Zak phase [17,19,20]. The results are $-\pi/2$ and $\pi/2$ for $\delta w > 0$ and $\delta w < 0$. Note that the Zak phase of each dimerization is a gauge dependent value, but the difference ($\Delta\varphi_{Zak} = \varphi_{Zak2} - \varphi_{Zak1} = \pi$) between Zak phases is topologically defined [20]. Therefore, the two dimerizations in Fig. 1(b) (red and blue) are topologically distinct to each other.

The distinct topologies ensure the existence of interface states between two connected dimerized chains with different Zak phases. Hereafter we cut and connect the two chains through their mirror centers. In the left part of Fig. 1(c), we apply $\delta w = 0.3w$ and $\delta w = -0.3w$ to the two connected chains, respectively. An interface state (red line), whose acoustic field is shown in Fig. 1(d), locates inside the bandgap. When $\delta w = 0.3w$ and $\delta w = 0.2w$ are applied to the two connected chains, no interface state is observed [right part of Fig. 1(c)].

We then construct 2D Dirac nodes by stacking these 1D dimerized chains. First, we consider the following 2D Bloch Hamiltonian:

$$H_2(k) = [2t_x \cos(k_x a) + 2t_y \cos(k_y a)]\sigma_x - 2\delta t_x \sin(k_x a)\sigma_y \qquad (2)$$

where $t_x$ ($t_y$) is the hopping strength along $x$ ($y$) direction, and $\delta t_x$ is the modulation of the hopping strength along the $x$ direction. After calculation, we find that there are two linear degenerate points in the first Brillouin zone (BZ) if $|t_x| < |t_y|$, no degenerate points if $|t_x| > |t_y|$, and quadratic degenerate points [21] in the corners of 2D BZ if $|t_x| = |t_y|$. Here we choose $t_x = -1$, $t_y = -2$ and $\delta t_x = -0.5$. The 2D band structure in Fig. 2(c) shows two isolated Dirac nodes that locate at $(0, \pm 2/3)$ in the first BZ [Fig. 2(b)].

Second, we design the unit-cell of the 2D acoustic lattice as shown in Fig. 2(a). The inset is the schematic of the lattice whose unit-cell is enclosed by green dashed lines. The lattice constant and parameters of the resonator are the same with those in Fig. 1(a). The modulation $\delta w_x = 0.3 w_x$ with $w_x = 0.26r$ is applied to coupling waveguides along the $x$ direction. Coupling waveguides along the $y$ direction with radius $w_y = 2w_x$ connect these staggered dimerized chains. The acoustic band structure along high symmetry lines in the first BZ in Fig. 2(d) shows two degenerate points with frequency $0.21 \times 2\pi c/a$ (718.05 Hz) where $c$ is the speed of sound, that locate at $(0, \pm 0.61)$.

These Dirac nodes can be regarded as topological charges characterized by winding number [22]. We find $w = -1$ ($w = 1$) for the Dirac node in $M_2\Gamma$ ($\Gamma M_3$) as shown in Fig. 2(d). The vortex structure (not shown here) of the normalized coefficients of Pauli matrix from Eqn. (2) can manifest the topological feature of these nodal points.

The 2D acoustic lattice has flat edge states similar to those in graphene [23]. We investigate the 2D structure that is finite in the $x$-$y$ direction with 15.5 unit-cells (even number of sites) and infinite in the $x+y$ direction, as plotted in Fig. 2(f). The band structure of this finite system is shown in Fig. 2(e). The red curves with degeneracy 2 indicate the nearly flat edge states connecting two projected Dirac nodes with opposite chirality. Note that the little derivation from a perfectly flat dispersion is a result of the real acoustic structure and hard boundary conditions adopted. The acoustic fields are shown in Fig. 2(f). The acoustic waves of degenerate edge states do not propagate due to nearly zero group velocity. Note that when the finite ribbon consists of 16 unit cells (odd number of sites), there will be a single edge state with degeneracy 1 that traverses the BZ, which is not discussed here.

Finally, by stacking the 2D dimerized lattice along the $z$ direction with periodicity $a$ and tuning the coupling strength, we can construct a Bloch Hamiltonian for the 3D dimerized lattice:

$$H_3(k) = d_0 I + d_x \sigma_x + d_y \sigma_y + d_z \sigma_z, \qquad (3)$$

where $d_x = 2t_x \cos(k_x a) + 2t_y \cos(k_y a)$, $d_y = -2\delta t_x \sin(k_x a)$, $d_z = t_{z1} \cos(k_z a) - t_{z2} \cos(k_z a)$, $d_0 = t_{z1} \cos(k_z a) + t_{z2} \cos(k_z a)$ and $I$ is the $2 \times 2$ identity matrix. The parameter $t_x$ ($t_y, t_z$) is the hopping strength along $x$ ($y$, $z$) direction. Note that the first term in Eqn. (3) plays the role of tilting the cone-like spectrum. With a strongly tilted cone spectrum, this Hamiltonian satisfies the condition of recently proposed type-II Weyl Hamiltonian [9]. Here we choose $t_x = -1$, $t_y = -2$, $t_{z1} = -1$, $t_{z2} = -2$ and $\delta t_x = -0.5$. The Hamiltonian breaks $P$-symmetry and respects $T$-symmetry. We calculate the band structure as shown in Fig. 3(b) with $k_z = 0.5$ in $k_x$-$k_y$ plane. Four linear degenerate points locate at $(0, \pm 2/3, \pm 1/2)$ in the 3D first BZ, which are also marked in Fig. 4(a). Typically, we plot in Fig. 3(c) the cone spectrum near the degenerate point $(0, 2/3, 1/2)$ in $k_y$-$k_z$ plane. It can be seen that the cone spectrum indeed has been strongly tilted. Since the group velocities near the degenerate point are $2t_{z1} \sin k_{z0}$ and $2t_{z2} \sin k_{z0}$ where $k_{z0}$ is the location of degenerate point, the two bands acquire the same sign of group velocity.

Figure 3(a) shows one unit-cell of the 3D acoustic structure. The inset presents the schematic of the 3D lattice. The radii of the coupling waveguides along the $z$ direction are $w_{z1} = w_x + 2\delta w_x$ for the A resonator and $w_{z2} = w_x$ for the B resonator. The other parameters are the same as in Fig. 2(a). The band structures in BZ planes $(k_x, k_y, 0)$ and $(k_x, k_y, 0.51)$ are shown in Fig. 3(d-e), which reveal a band gap at $k_z = 0$, and two degenerate points with frequency $0.26 \times 2\pi c/a$ (900.10 Hz) at $k_z = 0.51$. In the left part of Fig. 3(f), by sweeping $k_z$ at $(k_x, k_y) = (0, 0.61)$, we get the band structure with a degenerate point at $k_z = 0.51$. It can be seen that the two bands have the same sign of group velocity. Therefore, there are four acoustic type-II Weyl nodes that locate at $(0, \pm 0.61, \pm 0.51)$. One significant distinction between the type-I and type-II Weyl nodes appears in the DOS [9]. For type-I Weyl nodes, the DOS vanishes at the frequency of Weyl nodes. However, the DOS acquires finite values for type-II Weyl nodes due to the presence of unbounded two-band pockets. We retrieve the parameters from fitting the band data with the Eqn. (3) and plot the DOS that arises due to the type-II Weyl node in the right part of Fig. 3(f). The contribution of the rest iso-frequency surface to the DOS is not included. The peak indicates the location of the type-II Weyl node.

In the 3D momentum space, the Weyl nodes are topological monopoles of quantized Berry flux characterized by chirality [26] indicated by black "+" (green "-") in Fig. 4(a) or Chern number [24,25]. We adopt the method from Ref. 9 and calculate the Berry phases of the two bands over a closed sphere centered at a 3D degenerate node (0,2/3,1/2) as a function of polar angle $\theta$ (0 to $\pi$). As shown in Fig. 4(b), the Berry phase of the lower (upper) band changes from 0 to $2\pi$ ($2\pi$ to 0), which verifies that the degenerate node is a Weyl node with Chern number 1. The right inset show the Berry curvature around the Weyl node. The same calculation can be applied to identifying the charges of other Weyl nodes.

The nonzero Chern numbers imply the existence of topological surface states. We investigate the 3D acoustic structure that is finite in the *x-y* direction and infinite in the *x+y* and *z* directions. In this case, the Weyl nodes are projected along the $\vec{k}_x - \vec{k}_y$ direction, indicated by black and green dots in Fig. 4(f), with good quantum numbers $\vec{k}_{//} = (\vec{k}_x + \vec{k}_y)/\sqrt{2}$ and $\vec{k}_z$. Figure 4(c-d) show the projected band structures with fixed $\vec{k}_{//} = \pi/\sqrt{2}a$ [Fig. 4(c)] and $\vec{k}_{//} = 0$ [Fig. 4(d)], as indicated by "Cut 1" and "Cut 2" in Fig. 4(f). In Fig. 4(c), two surface states (red and green curves), corresponding to the two opposite surfaces, locate in an incomplete bandgap and both acquire positive group velocity. In Fig. 4(d), no surface states show up. As presented in Fig. 4(e), the upper (lower) acoustic field corresponds to the surface state of green (red) curve in Fig. 4(c). We also plot the sound intensity $I = pv$ ($p$ is sound pressure and $v$ is the velocity) as grey arrows, whose length represents the amplitude of sound intensity. Both the two surface states propagate along the *z* direction, which is consistent with the positive group velocity in Fig. 4(c).

Note that the positive group velocity of the surface states is determined by the tilting direction of type-II Weyl nodes. In contrast, in the previously demonstrated topological surface states [8] of acoustic type-I system, their propagation direction is surface-dependent: if the surface states on one surface propagate in one direction, those on the opposite surface should propagate in the opposite direction. This distinction is schematically illustrated in Fig. S1 [26]. Another distinction is that the surface states of type-II Weyl nodes stay in an incomplete bandgap. With a single-frequency excitation (simulated results in Fig. S2 [26]), the surface states will be scattered by defects and penetrate into the bulk. Therefore, they do not have the same robustness as demonstrated in Ref. 8, but the existence of "Fermi arcs" connecting Weyl nodes is still topologically protected.

To demonstrate the acoustic "Fermi arcs," we trace out the trajectories of surface states at frequency 0.26×2π*c/a* (fixed frequency plays the role of Fermi energy) in the 2D BZ $(\vec{k}_{//}, \vec{k}_z)$, as indicated by red dotted points in Fig. 4(f). Here we consider a semi-infinite system and thus only the surface states localized at one surface [red curve in Fig. 4(c)] are included. These trajectories indeed connect two pairs of type-II Weyl nodes, as an analog of Fermi arcs in type-II Weyl semimetals [9-13].

The above results demonstrate the feasibility of constructing acoustic type-II Weyl nodes by stacking 1D dimerized chains of acoustic resonators. The unique features of acoustic type-II Weyl system, such as the finite DOS and transport properties of surface states, are demonstrated. The Fermi-arc-like surface states can be traced out as an analog of Fermi arcs in recently demonstrated type-II Weyl semimetals. The stacking method shown in this work provides an approach of constructing topological phases at different dimensions with the same building blocks, and may be extended to other systems including cold atoms [20,27], photonics [28-33], and polaritons [34-36].

The authors thank L. Lu, Y. D. Chong, A. A. Soluyanov, M. Xiao and F. Gao for helpful discussions. This work was sponsored by Nanyang Technological University under Start-Up Grants, and Singapore Ministry of Education under Grant No. MOE2015-T2-1-070 and Grant No. MOE2011-T3-1-005.

**References**
1. H. Weyl. I. Z. Phys. 56, 330–352 (1929).
2. X. Wan, et al. Phys. Rev. B 83, 205101 (2011).
3. S.-Y. Xu, et al. Science 349, 613–617 (2015).
4. B. Q. Lv, et al. Phys. Rev. X 5, 031013 (2015).
5. L. Lu, et al. Nat. Photonics 7, 294–299 (2013).
6. L. Lu, et al. Science 349, 622–624 (2015).
7. L. Lu, J. D. Joannopoulos and M. Soljacic. Nat. Photonics 8, 821-829 (2014).
8. M. Xiao, W. Chen, W. He and C. T. Chan. Nat. Physics 11, 920-924 (2015).
9. A. Soluyanov, et al. Nature 527, 495-498 (2015).
10. S. Xu, et al. arXiv: 1603.07318 (2016).
11. J. Jiang, et al. arXiv: 1604.00139 (2016).
12. A. Liang, et al. arXiv: 1604.01706 (2016).
13. N. Xu, et al. arXiv: 1604.02116 (2016).
14. Z. Yang, et al. Phys. Rev. Lett. 114, 114301 (2015).
15. X. Ni et al. New J. Phys. 17, 053016 (2015).
16. B. Khanikaev et al. Nat. Commun. 6, 8260 (2015).
17. M. Xiao, et al. Nat. Physics 11, 240-244 (2015).
18. W. P. Su, et al. Phys. Rev. Lett. 42, 1698-1701 (1979).
19. J. Zak, Phys. Rev. Lett. 62, 2747-2750 (1989).
20. M. Atala, et al. Nat. Physics 9, 795-800 (2013).
21. Y. D. Chong, X. G. Wen and M. Soljacic. Phys. Rev. B 77, 235125 (2008).
22. K. Sun, W.V. Liu, A. Hemmerich and S. Das Sarma. Nat. Physics 8, 67 (2011).
23. A. Castro Neto, et al. Rev. Mod. Phys. 81, 109–162 (2009).
24. M. Z. Hasan, and C. L. Kane, Rev. Mod. Phys. 82, 3045–3067 (2010).
25. X. L. Qi, and S. C. Zhang, Rev. Mod. Phys. 83, 1057-1110 (2011).
26. Supplementary material online, which includes Ref. [37].
27. S. Ganeshan and S. Das Sarma. Phys. Rev. B 91, 125438 (2015).
28. Z. Wang, et al. Nature 461, 772-775 (2009).
29. M. C. Rechtsman et al. Nature 496, 196-200 (2013).
30. M. Hafezi, et al. Nat. Photonics 7, 1001-1005 (2013).
31. F. Gao, et al. Nat. Commun. 7, 11619 (2016).
32. W. Chen, et al. Nat. Commun. 7, 13038 (2016).
33. M. Xiao, Q. Lin and S. Fan. Phys. Rev. Lett. 117, 057401 (2016).
34. A. V. Nalitov, et al. Phys. Rev. Lett. 114, 026803 (2015).
35. A. V. Nalitov, et al. Phys. Rev. Lett. 114, 116401 (2015).
36. T. Karzig, et al. Phys. Rev. X 5, 031001 (2015).
37. M. Hafezi, et al, Nature Phys. 7, 907-912 (2011).

**Figure legends**

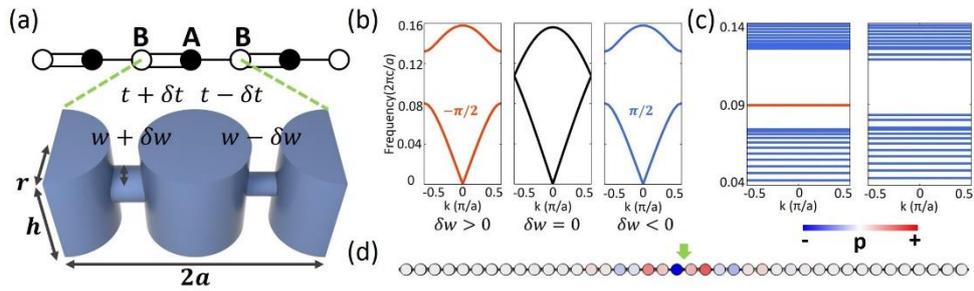

**Figure 1. Dimerized chain of acoustic resonators.** (a) The schematic of a dimerized chain and one unit-cell of the acoustic structure. (b) Three band structures with $\delta w = 0.3w$, $\delta w = 0$ and $\delta w = -0.3w$. (c) Topologically non-trivial and trivial band structures for the interface between two connected chains. The red line indicates the topological interface state. (d) The acoustic field of the interface state. The green arrow points to the interface. Hereafter, the blue (red) color represents negative (positive) acoustic pressure (*p*) in the colorbar.

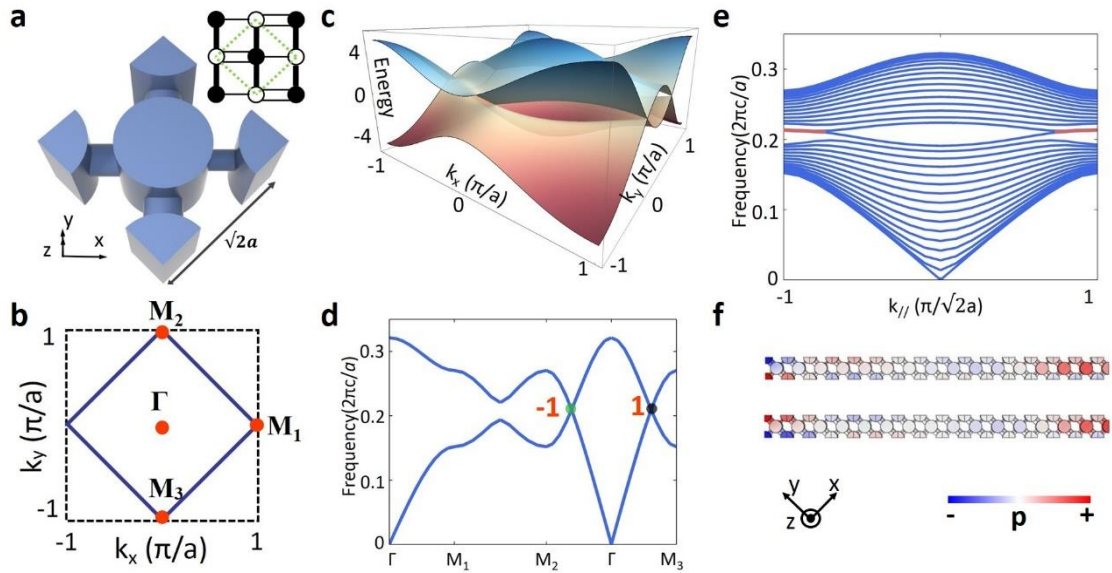

**Figure 2. 2D acoustic dimerized lattice.** (a) The schematic of the 2D dimerized lattice and one unit-cell of the acoustic structure. (b) The first BZ enclosed by the blue lines. (c) The band structure from the tight-binding model. (d) The band structure of the acoustic lattice. Red numbers indicate the chirality. (e) The band structure for a finite acoustic structure. The red curves indicate the flat edge states. (f) The acoustic fields of the edge states.

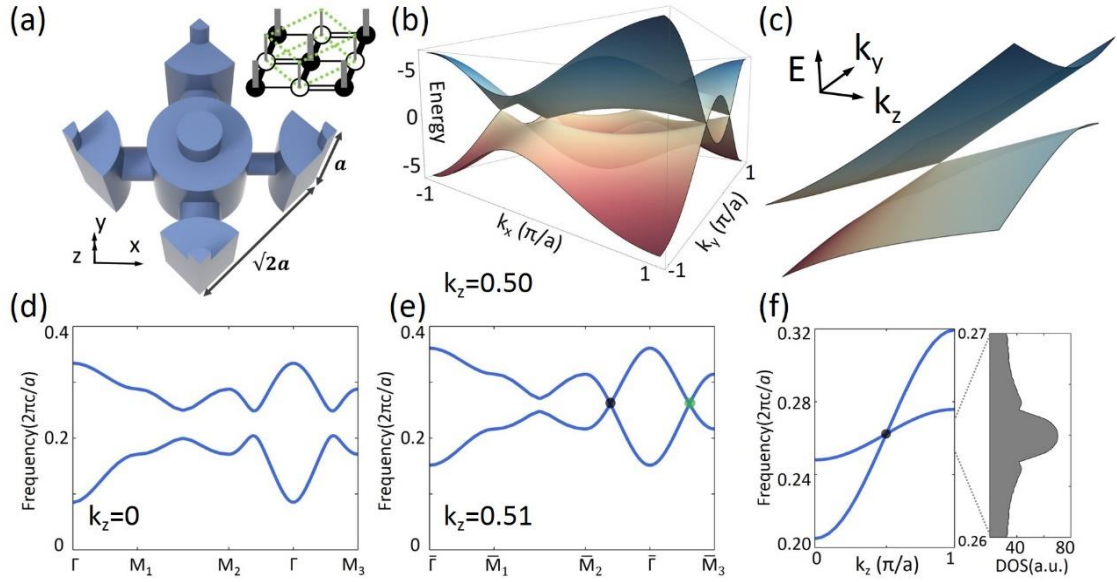

**Figure 3. 3D acoustic type-II Weyl nodes.** (a) The schematic of the 3D dimerized lattice and one unit-cell of the acoustic structure. (b-c) The tight-binding band structures in the BZ plane with $k_z = \pi/2a$ (b) and around the Weyl node with $k_x = 0$ (c). (d-e) The band structures along the high symmetry lines in 2D BZ planes. The black (green) dot indicates the Weyl node with positive (negative) chirality. (f) Left: the band structure with fixed $(k_x, k_y) = (0, 0.6l)$. Right: density of states.

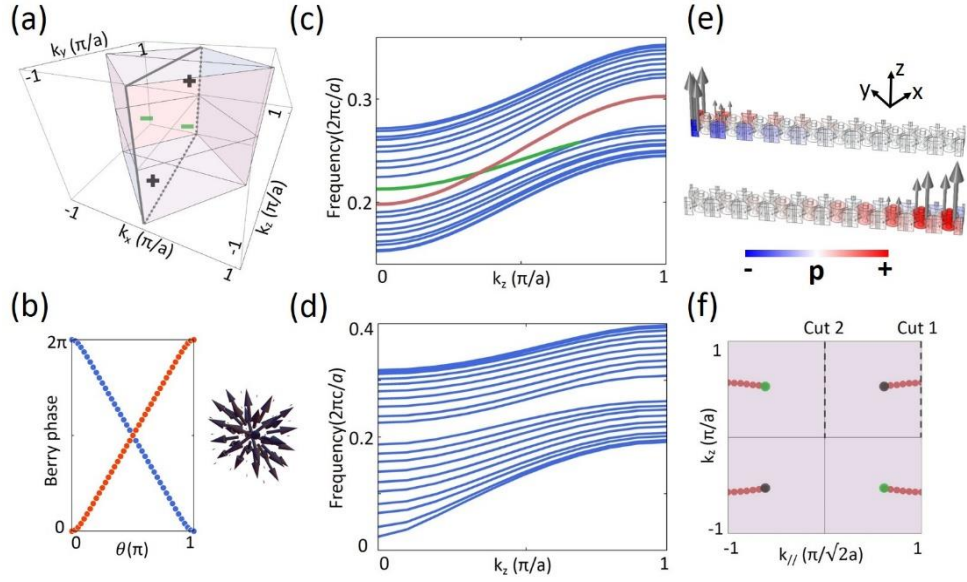

**Figure 4. Chirality and Fermi arc-like surfaces states.** (a) Type-II Weyl nodes in 3D first BZ. (b) Berry phase and Berry curvature around the Weyl point. (c-d) The band structures with $\vec{k}_{//} = \pi/\sqrt{2}a$ (c) and $\vec{k}_{//} = 0$ (d). The red and green curves indicate the surface states. (e) The acoustic fields of the surface states in panel (c). The grey arrows represent the sound intensity. (f) Red dotted points indicate the trajectories of the "Fermi Arc".